\documentclass[aps,pra,longbibliography,twocolumn]{revtex4-1} 

\usepackage{graphicx}
\usepackage[hidelinks]{hyperref}

\usepackage[percent]{overpic} 
\usepackage{ulem} 
\usepackage{xcolor} 
\usepackage{mdframed} 
\newmdenv[backgroundcolor=green!10!white, linecolor=white, leftmargin=10pt, innerleftmargin=5pt,innertopmargin=5pt,innerrightmargin=5pt,innerbottommargin=5pt]{answer}

\newcommand{\unit}[1]{\,\mathrm{#1}}

\begin{document}
    
    \title{Striped states in a many-body system of tilted dipoles}
    
    
    \author{Matthias Wenzel}
    \author{Fabian B\"ottcher}
    \author{Tim Langen}
    \author{Igor Ferrier-Barbut}
    \author{Tilman Pfau}
    \email{t.pfau@physik.uni-stuttgart.de}
    
    \affiliation{5{.} Physikalisches Institut and Center for Integrated Quantum Science and Technology, Universit{\"a}t Stuttgart, Pfaffenwaldring 57, 70569 Stuttgart, Germany}

    \date{\today}
    
    \begin{abstract}
        We study theoretically and experimentally the behaviour of a strongly confined dipolar Bose-Einstein condensate, in the regime of quantum-mechanical stabilization by beyond-mean-field effects. Theoretically, we demonstrate that self-organized ``striped'' ground states are predicted in the framework of the extended Gross-Pitaevskii theory. Experimentally, by tilting the magnetic dipoles we show that self-organized striped states can be generated, likely in their metastable state. Matter-wave interference experiments with multiple stripes show that there is no long-range off-diagonal order (global phase coherence). We outline a parameter range where global phase coherence could be established, thus paving the way towards the observation of supersolid states in this system.
    \end{abstract}
    
    
    
    \maketitle

    Self-organization is a key many-body phenomenon, in which order arises spontaneously due to interactions between the individual constituents. An extraordinary example in the quantum world are supersolid states of matter, which feature a self-organized density-modulation, but unlike an ordinary crystal, also maintain superfluid properties.
    Amongst others, possible candidates are ultracold atoms with a non-negligible dipole-dipole interaction. The roton minimum \cite{Santos2003} in the dispersion relation of these dipolar Bose-Einstein condensates (BECs) introduces a natural length scale for self-organization of the system. However, the density-modulated states induced by the roton softening are unstable within mean-field theory \cite{Komineas2007}. In contrast to this prediction, stable liquid-like ``quantum droplets'' were observed \cite{Kadau2016} with dysprosium atoms. The initial observation was followed by the identification of the underlying stabilization mechanism \cite{Ferrier-Barbut2016} and studies of the collective oscillations with erbium atoms \cite{Chomaz2016a} as well as the observation of self-bound single droplets \cite{Schmitt2016a} of dysprosium.
    With a theoretical description of the single-droplet physics at hand \cite{Wachtler2016a,Baillie2016}, here we turn to the fundamental question of supersolidity in systems of multiple quantum droplets. The density-modulation of these states would break the translational symmetry. In addition, their global common superfluid phase breaks an internal gauge symmetry. In contrast to recent observations of supersolidity induced by an optical lattice in a double cavity \cite{Leonard2017} and a spin-orbit-coupled BEC \cite{Li2017} where the period of the density modulation is imprinted by an external light field, the density modulation of this system would thus be exclusively due to the intrinsic anisotropic long-range dipole-dipole interaction of the atoms.
    
    \begin{figure}[ht]
            \begin{overpic}[width=0.45\textwidth]{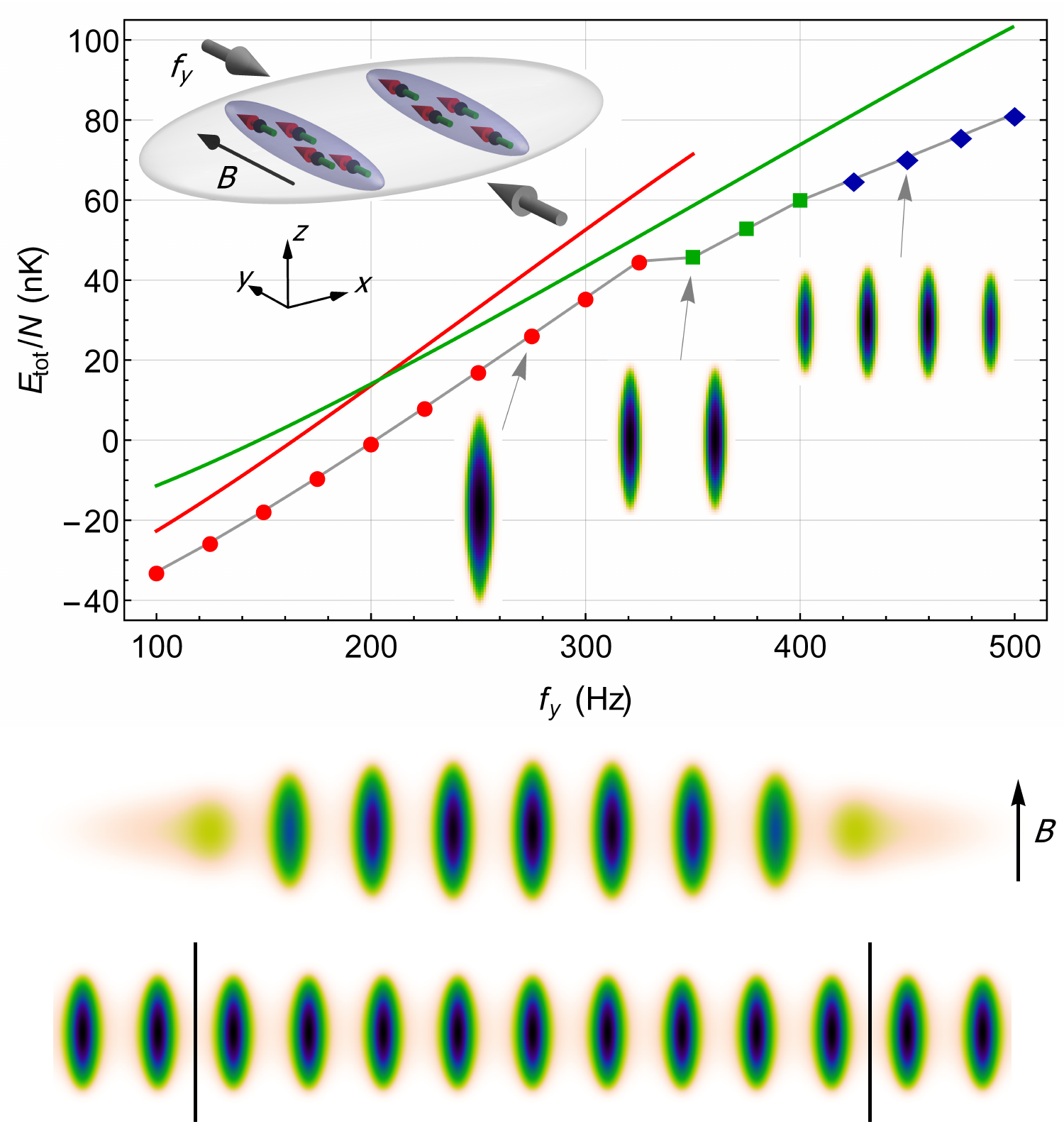}
            \put(0,94){a)} \put(0,30){b)} \put(0,14){c)}\end{overpic}
        \vspace{-3mm}
        \caption{\label{fig:1} Striped ground states.
            a) Total energy per atom $E_\mathrm{tot}/N$ for single (solid red) and double (solid green) droplet solutions obtained from a variational ansatz. For $f_y \gtrsim 200\unit{Hz}$ the state with two droplets has lower energy than a state with a single droplet. The schematic shows a double droplet configuration in the proposed harmonic trap.
            Numerical simulations of the extended Gross-Pitaevskii equation, see eq{.} (\ref{eq:eGPE}), predict higher numbers of droplets for increasing $f_y$ in the ground state. Insets show the integrated column density along $z$ for ground states with one (red dots), two (green squares) and four (blue diamonds) droplets.
            b) Integrated column density of the ground state for $f_y = 800\unit{Hz}$. There are several droplets with finite density between the droplets indicating overlap of the single droplet wavefunctions.
            c) For a similar system with periodic boundary conditions along $x$ ($f_x = 0$) the ground state exhibits the same density modulation, thus breaking the continuous translational symmetry. Vertical lines represent the edge of the box. See text for further parameters.
        }
    \end{figure}
        
    In the following we first demonstrate theoretically that the ground state of a dipolar Bose gas in an anisotropic harmonic trap is a ``striped state'' featuring multiple droplets for a certain parameter range.
    This extends prior theoretical work predicting only single-droplet ground states \cite{Wachtler2016} within the framework of the extended Gross-Pitaevskii (eGPE) equation \cite{Wachtler2016a,Baillie2016}. All experiments observing multiple droplets \cite{Kadau2016,Ferrier-Barbut2016,Ferrier-Barbut2016a} have been carried out in weak traps where the ground state is a single droplet.
    
    Striped phases and states are well known and studied in electronic systems in superconducting materials \cite{Orenstein2000} and have been predicted for dipoles in two dimensions \cite{Baranov2012}. In this paper we refer to striped states in confined geometries. We term the found states this way in order to express the fact that the system as a whole forms a collectively ordered ground state, where anisotropy plays a crucial role. More precisely, we present a symmetry-breaking effect along an axis perpendicular to the confinement.
    
    Second, we experimentally realize and study such an ensemble of dipoles in a constrained geometry. By tilting the polarizing magnetic field we effectively tune the mean-field dipolar interaction and are able to control the number of droplets with the underlying trap. Using these novel tools we observe striped states with higher droplet numbers than expected from theory and discuss their nature. We further conduct expansion measurements of these states to investigate the coherence properties and outline a way to reach phase coherence of the whole system.

    \section{Theory}

    Dipolar quantum droplets exist thanks to the interplay between attractive mean-field interactions and a repulsive beyond mean-field correction arising from quantum fluctuations \cite{Ferrier-Barbut2016,Chomaz2016a,Schmitt2016a}. These states are liquid-like, featuring a low compressibility while  preserving a peak density that is an order of magnitude higher compared to the condensate phase.
    In contrast to typical liquids, the binding mechanism of the liquid relies on the dipoles being mainly in a head-to-tail configuration, and thus on an anisotropic density distribution.
    
    When compressing a usual liquid droplet along one or two directions, the droplet changes its overall shape to conserve its volume and thus its density. In the case of dipolar quantum droplets, the shape cannot be strongly altered without breaking the binding mechanism. Thus, a strong confinement in two directions including the droplet long axis (magnetic field axis), will lead to a strong frustration, since both the nearly constant peak density and the anisotropy cannot be kept simultaneously. Therefore one expects a different ground state in a strongly constrained geometry. For this reason, states with multiple droplets might have lower energy than the single-droplet states.
    In fact quantum Monte Carlo calculations predict ground states with multiple dipolar droplets for very low atom numbers \cite{Macia2016,Cinti2017}. However, these calculations feature molecular potentials that do not contain the short-range (van der Waals) interaction. As recently shown \cite{Odziejewski2016} the realistic scattering potential is well described by including contact and dipolar interactions as practiced throughout this article.
    
    To investigate the behaviour of quantum droplets in a constrained geometry we use the framework of the extended Gross-Pitaevskii equation (eGPE) which was verified by quantum Monte Carlo simulations \cite{Saito2016} and describes well recent experiments with dipolar droplets$\ $\cite{Schmitt2016a,Chomaz2016a}. The eGPE itself
    \begin{equation}\label{eq:eGPE}
    i\hbar \partial_t \psi = \bigg[ -\frac{\hbar^2 \nabla^2}{2m} + V_\mathrm{ext} + g |\psi|^2 + \Phi_\mathrm{dip} + g_{qf} |\psi|^3 \bigg] \psi
    \end{equation}
    includes an external potential $V_\mathrm{ext} = \sum_k\frac{1}{2}m\omega_k^2 k^2$ ($\omega_k = 2\pi f_k$ with $k=x,y,z$), the mean-field contact, and dipolar interaction potential
    \begin{equation}\label{eq:Phidd}
    \Phi_\mathrm{dip}(\mathbf{r}) = \frac{3 \, g_{dd}}{4\pi}  \int\!\mathrm{d}\mathbf{r}^\prime \frac{1-3\cos^2(\vartheta)}{|\mathbf{r}-\mathbf{r}^\prime|^3} |\psi(\mathbf{r}^\prime)|^2
    \end{equation}
    with $\vartheta$ being the angle between the polarization direction of the dipoles and their relative orientation \cite{Lahaye2009}, and a term $g_{qf} |\psi|^3$ with $g_{qf} = \frac{32 \,g \,a_s^{3/2}}{3 \,\pi^{1/2}} \big(1 + \frac{3}{2} \frac{a_{dd}^2}{a_s^2}\big)$ taking into account quantum fluctuations within a local density approximation \cite{Lima2011}.
    The main assumptions of this zero-temperature model are thus the validity of the local density approximation, the weakness of the quantum depletion and the validity of the interaction potential resulting from the first-order Born approximation \footnote{The validity of the Born approximation has been investigated for Dy in \cite{Odziejewski2016}. In our temperature range we expect an effective shift of the dipole length of a few percent which is beyond the accuracy of this work.}.
    In the following, we consider $N = 10^4$ $^{164}\mathrm{Dy}$ atoms with contact and dipolar interaction in an anisotropic harmonic trap. The contact interaction is defined by the scattering length $a_s = 70\,a_0$ with $g = 4 \pi \hbar^2 a_s / m$. The reported values for the background scattering length of Dy are somewhat higher \cite{Tang2015a,Maier2015}, however the critical atom number for self-bound states observed in our earlier work implies a background value of $a_{bg} \approx 63\,a_0$. The exact value is still uncertain, so we pick an intermediate value. The dipolar length $a_{dd} = 131\,a_0$ with the corresponding $g_{dd}$ is a result of the strong magnetic moment $\mu = 9.93\,\mu_B$ of Dy. We further assume a magnetic field $B \parallel \hat{y}$ and an anisotropic trapping potential with $f_x = 70\unit{Hz}$, $f_y = 10 - 500\unit{Hz}$ and $f_z = 1\unit{kHz}$ (see the schematic inset in Fig{.} \ref{fig:1}a).
    
    To gain initial insight, we perform semi-analytical calculations, making use of a variational ansatz. As detailed in \cite{SupMat} we extend the variational ansatz of a Gaussian density profile \cite{Wachtler2016a,Baillie2016} to a double-droplet state consisting of two droplets with Gaussian wavefunctions characterized by sizes $\sigma_x,\sigma_y,\sigma_z$ and inter-droplet distance $d$.
    By minimizing the energy with respect to these parameters using the energy functional corresponding to eq. (\ref{eq:eGPE}),  we can determine the ground state within this ansatz. As a function of $d$, we find in general two energy minima \cite{SupMat}. One at $d=0$ (solid red) corresponding to a single droplet and one at $d > \sigma_x > 0$ (solid green) corresponding to two separated droplets, as presented in Fig{.} \ref{fig:1}a). For increasing $f_y$ the ground state changes from the single-droplet to the double-droplet state.
    This confirms the scenario outlined earlier whereby the combined effect of liquid-like properties and the strong anisotropy leads to a density-modulated ground state.
    
    Further increasing the confinement is thus expected to favor states with higher droplet numbers, which are not covered by the variational approach. To fully confirm our scenario within the eGPE and to account for the possibility of more than two droplets we perform, as in$\ $\cite{Schmitt2016a}, full numerical simulations of the eGPE to find the ground state of the system via imaginary time evolution.
    Within the investigated parameter range, we indeed obtain ground states with a single (red dots), two (green squares) or even four droplets (blue diamonds). Fig{.} \ref{fig:1}a) shows the total energy per atom and the number of droplets with representative insets of the column density integrated along $z$.
    In general, higher confinement along the droplet axis (see Fig{.} \ref{fig:1}b) as well as increasing the total atom number leads to a larger number of droplets in the ground state. For even larger confinement along $y$ the ground state becomes a ``uniform'' gaseous BEC phase. We note that compared to the variational ansatz the frequency $f_y$ at which the ground state splits up into two droplets is shifted to a slightly higher value.
    
    To verify that this is a general effect independent of the $x$ confinement, we study the case of an infinite system along $x$. We find that the system exhibits the same transition to a density-modulated ground state, as can be seen in Fig{.} \ref{fig:1}c). This state is realized with periodic boundary conditions, a linear density $n_x = 800\unit{\mu m^{-1}}$ as well as trap frequencies $f_x = 0$, $f_y = 800\unit{Hz}$ and $f_z = 1000\unit{Hz}$. In this case the system features a continuous translational symmetry, that is broken by the transition to a periodic density-modulated ground state. The corresponding length-scale is determined by the interplay between $y$ confinement and interactions.
    
    Although multiple droplets have been observed in prior experiments \cite{Kadau2016,Ferrier-Barbut2016} these have been carried out with weak trapping potentials. In the following section we present our experimental advances towards the theoretically predicted multi-droplet regime.

    \section{Experiment}

    In order to explore the behaviour of strongly dipolar bosons in confined geometries like the ones considered above, we extended our experimental apparatus presented in \cite{Schmitt2016a} in two ways. First, we use three pairs of coils that allow us to orientate the magnetic field along an arbitrary direction. Second, we implement a light sheet that strongly confines the atoms along the vertical $z$ direction, which allows us to realize the strongly anisotropic traps that we studied in the theoretical analysis. Details are given in \cite{SupMat}.
    We always start our experimental sequence with a BEC of $^{164}$Dy containing about 5000 atoms at a temperature of $T\approx30\unit{nK}$ ($\approx 30\%$ thermal fraction) and the magnetic field along $z$. The trapping potential is oblate with trapping frequencies above $250\unit{Hz}$ along $z$. In this configuration the atomic dipoles arrange side-by-side, the dipole-dipole interaction (DDI) is repulsive, and we observe that the angular-roton instability we reported in \cite{Kadau2016} is prevented. Our ability to rotate the magnetic field in the plane of the light sheet allows us to effectively tune the DDI, turning it from repulsive to attractive. Experiments are carried out in the vicinity of a Feshbach resonance (position $B_0 = 1326(3)\unit{mG}$ and width $\Delta = 8(5)\unit{mG}$ \cite{Kadau2016}) to tune the scattering length.
    
    In a first set of experiments we study the behaviour of a BEC in an oblate cylindrical trap ($f_x \approx f_y \ll f_z$) with the magnetic field tilted in the plane. In a second set we investigate this system breaking the radial trap isotropy ($f_x \neq f_y$) allowing us to study geometries considered in the previous section.
    
    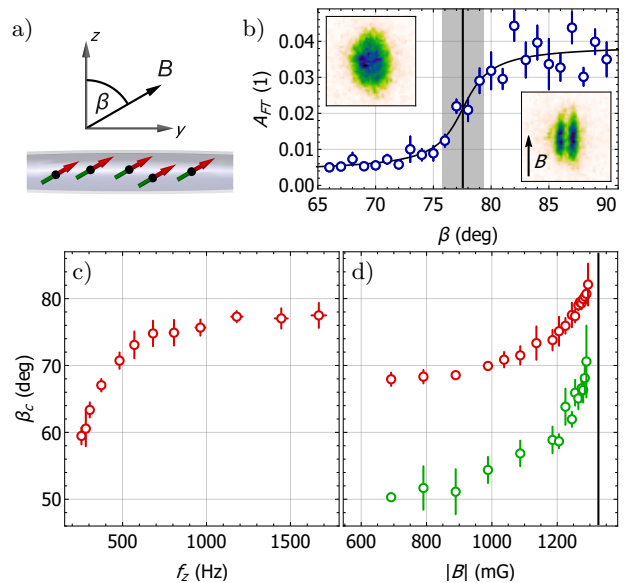
\begin{figure}
        \begin{overpic}[width=0.45\textwidth]{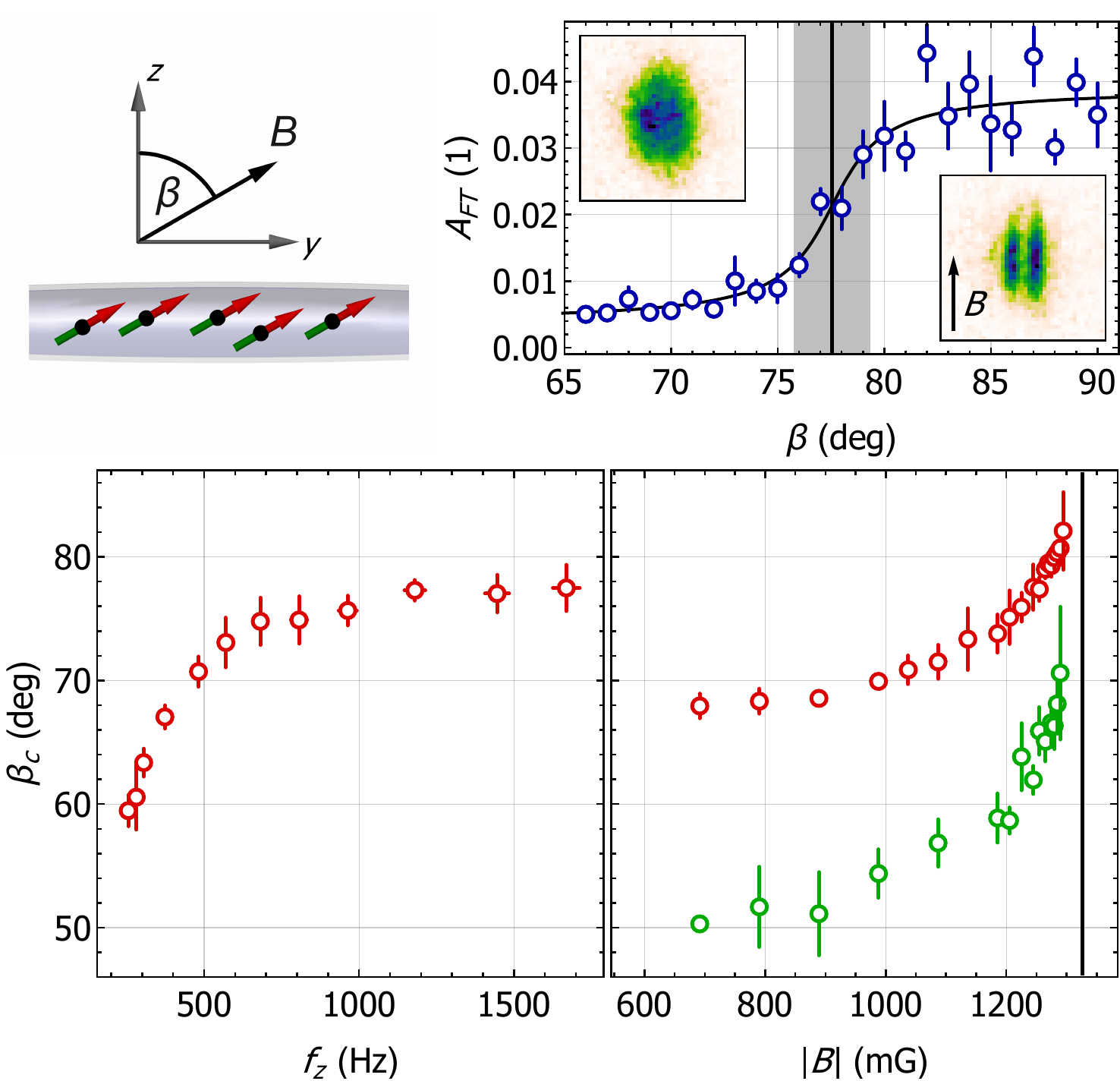}
            \put(0,90){a)} \put(38,90){b)} \put(10,50){c)} \put(56,50){d)}
        \end{overpic} \vspace{-3mm}
        \caption{\label{fig:2} Tuneability of the mean-field dipolar interaction $\Phi_\mathrm{dip}$.
            a) Schematic of dipolar atoms strongly confined along the $z$ direction with a magnetic field tilted under an angle $\beta$ with respect to the confinement axis.
            b) Determination of the critical angle $\beta_c$ via the Fourier anisotropy $A_{FT}$, see \cite{SupMat}. The gray bar marks the corresponding error, insets show a BEC elongated by magnetostriction (left) and a double droplet state (right).
            Measured critical tilt angle $\beta_c$ over c) $z$ trap frequency and d) magnetic field close to a Feshbach resonance.
            In c) the power of the light sheet is varied at fixed magnetic field $B = 1240(5)\,\text{mG}$.
            d) The dependence on the scattering length is measured by changing the magnetic field for $f_z = 950(10)\unit{Hz}$ (red) and $300(10)\unit{Hz}$ (green). The Feshbach resonance at $B_0 = 1326(3)\unit{mG}$ with width $\Delta = 8(5)\unit{mG}$ \cite{Kadau2016} is marked.}
    \end{figure}
    
    Let us describe our first set of experiments. In a trap with fixed frequencies, we tilt the magnetic field along the $y$ axis (angle $\beta$ with respect to $z$ axis, see Fig{.} \ref{fig:2}a),  at a constant amplitude $B$ and rate $\dot \beta = 0.33\unit{deg/ms}$. We find that the results of our experiments are independent of the field rate for rates slower than $\dot \beta \le 0.4\unit{deg/ms}$, see \cite{SupMat}.
    When tilting the magnetic field we first observe magnetostriction of the Bose-Einstein condensate: the elongation of the condensate along the magnetic field axis \cite{Lahaye2009}, reported here for the first time in-situ rather than during time-of-flight. Subsequently, we observe a sharp transition at an angle $\beta_c$ to a state consisting of one or two droplets, that are very elongated in the plane, as depicted in the inset of Fig{.} \ref{fig:2}b). The number of droplets varies between one and two depending mainly on the atom number.
    We quantify this transition by measuring the anisotropy $A_{FT}$ of the Fourier-transformed images, which we define in \cite{SupMat}. This linear image analysis allows us to avoid fitting the observed distribution with a simplified function, and provides a measure of the anisotropy both for one or several droplets. The Fourier anisotropy does exhibit a sharp step-like form, which allows us to extract the critical angle $\beta_c$, see Fig{.} \ref{fig:2}b).
    
    We measure the critical angle $\beta_c$ for varying $z$ trap frequency from $f_z = 255(15)$ to $1669(43)\unit{Hz}$. The trap is almost cylindrical with transversal trap frequencies $f_x = 46(1) - 53(2)\unit{Hz}$ and $f_y = 46(1) - 60(2)\unit{Hz}$ leading to a trap aspect ratio $\lambda = f_z / \sqrt{f_x f_y} = 5.5(4) - 29.6(8)$, while the magnetic field amplitude is fixed to $B = 1240(5)\unit{mG}$.
    As shown in Fig{.} \ref{fig:2}c) the critical tilt angle saturates for $f_z \gtrsim 900\unit{Hz}$ becoming independent of the confinement along the z axis.
    Next, we measure the critical angle for varying magnetic field amplitude approaching the Feshbach resonance. The trap is cylindrical with frequencies $f_z = 950(10)\unit{Hz}$ (red) and $300(10)\unit{Hz}$ (green) as well as $f_x = f_y = 50(5)$ and $48(5)\unit{Hz}$, respectively. According to $a_s(B) / a_{bg} = 1 + \Delta/(B_0-B)$ \cite{Chin2010} the scattering length $a_s$ is expected to vary between $1.01\,a_{bg}$ and $1.25\,a_{bg}$ in the measured range $B = 692(4) - 1294(4)\unit{mG}$.
    
    Both measurements demonstrate that the effective dipolar interaction $\Phi_\mathrm{dip}$ can be tuned by means of a tilted magnetic field. 
    Our experimental procedure thus offers a way to prepare the system in a given geometry, strongly confined in the $z$ direction as is necessary to observe the ground states predicted earlier, and then drive a transition with a continuous magnetic field tilt.
    
    \begin{figure} \vspace{-2mm}
        \begin{overpic}[width=0.45\textwidth]{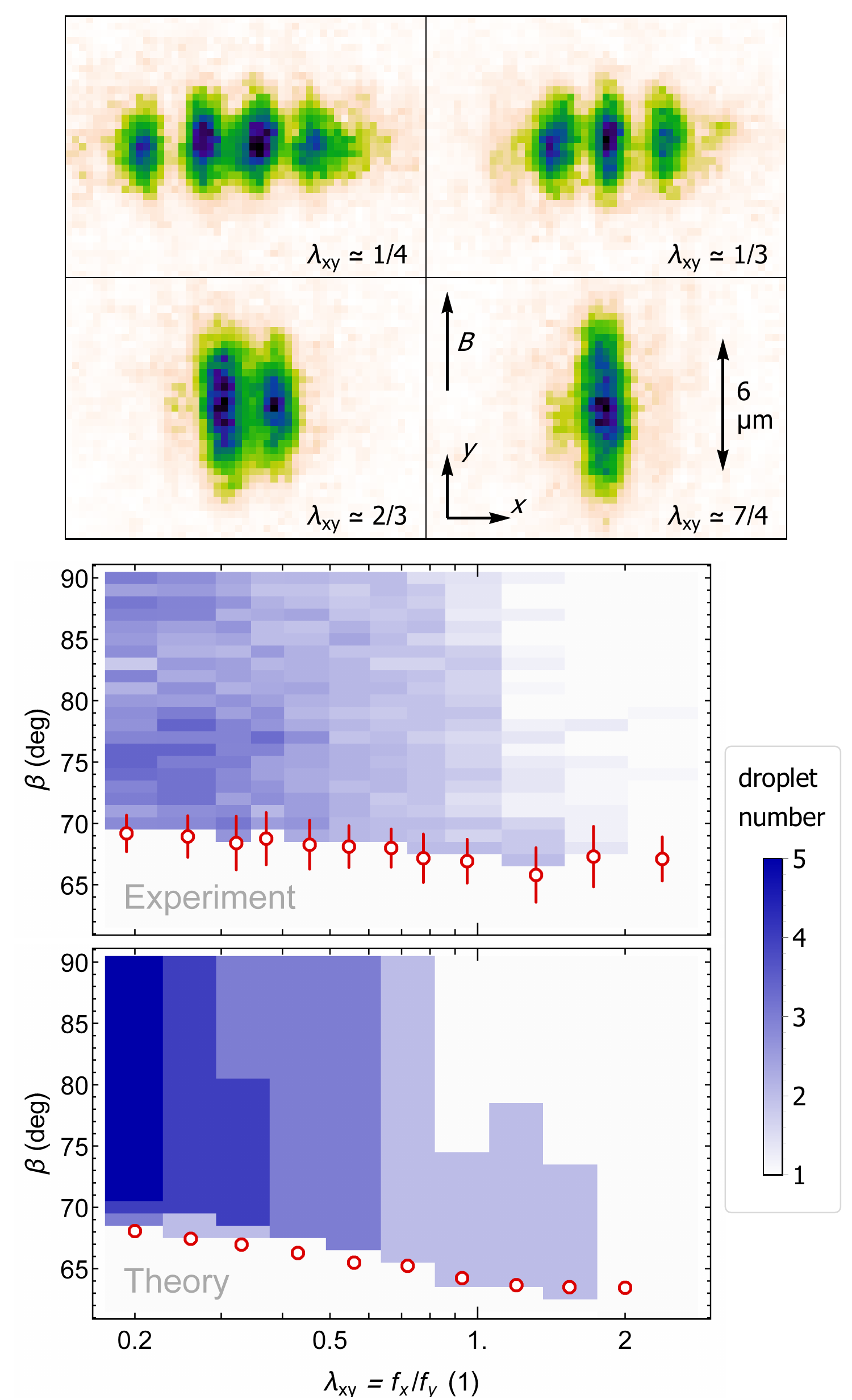}
            \put(1,96.5){a)} \put(1,56.5){b)} \put(1,29){c)}
        \end{overpic} \vspace{-2mm}
        \caption{\label{fig:3} Striped states observed in an anisotropic trap.
            a) Example single-shot \textit{in situ} images for varying transversal trap aspect ratio $\lambda_{xy} = f_x/f_y$.
            b) Critical tilt angle $\beta_c$ (red circles) and average number of droplets over $\lambda_{xy}$. We observe multiple droplets for $\lambda_{xy} \lesssim 1$ and single ones for $\lambda_{xy} \gtrsim 1$. Data is taken for $B=905(5)\unit{mG}$ at a trap frequency of $f_z = 945(5)\unit{Hz}$ and averaged over 11 realizations.
            c) Dynamic simulations of the eGPE confirm the creation of multiple droplets for conditions where a single droplet is the ground state. Simulation parameters are similar to the experiment, see main text and \cite{SupMat}.}
    \end{figure}

    The second set of experiments aims at determining whether density-modulated ground states can be observed experimentally. In order to do so we also reshape the trap in the $xy$-plane prior to tilting the magnetic field. We vary the transversal trap aspect ratio $\lambda_{xy} = f_x/f_y$ such that the mean trap frequency $\bar f = (f_x f_y f_z)^{1/3}$ is kept constant. For the range $\lambda_{xy} = 0.19 - 2.36$ the trap frequencies are varied in the range $f_x = 25(1) - 75(2) \unit{Hz}$ and $f_y = 128(2) - 32(1)\unit{Hz}$ with $f_z = 945(5)\unit{Hz}$ and the magnetic field amplitude $B = 906(5)\unit{mG}$ fixed. The atom numbers in the droplets are in the range $1000 - 3000$ with additional $\approx$ 6000 thermal atoms.
    While we obtain a single droplet for $\lambda_{xy} \gtrsim 1$ we find striped states of multiple droplets for $\lambda_{xy} \lesssim 1$. The in situ images presented in Fig{.} \ref{fig:3}a) are examples of single realizations of these states.
    For a quantitative analysis we vary the tilt angle $\beta$ for different $\lambda_{xy}$ values, see Fig{.} \ref{fig:3}b). While we observe a negligible change in the critical tilt angle $\beta_c$ (red dots) within error bars, the mean number of droplets varies. For $\lambda_{xy} > 1$ the trap is elongated along $\hat y$, the direction of the magnetic field tilt. In such a configuration with weaker confinement along the tilt axis we obtain a single droplet. In the opposite case where $\lambda_{xy} < 1$ the confinement along the tilt axis is stronger leading to a mean droplet number of up to $3.4$.
    In contrast to prior observations of quantum droplets \cite{Kadau2016,Ferrier-Barbut2016a} we gain control over the number of created droplets here.
    
    As shown in our theoretical analysis the number of droplets increases with increasing confinement $f_y$ (decreasing $\lambda_{xy}$) along the in-plane magnetic field component $B_y$, as summarized in Fig{.} \ref{fig:1}a). In addition, dynamic simulations presented in Fig{.} \ref{fig:3}c) show excellent agreement with the experiment in both the critical angle and the dependence of droplet number on $\lambda_{xy}$. The parameters for the simulation are $N = 5000$ atoms, scattering length $a_s = 70 \, a_0$, trapping potential as in the experiment and finite three-body losses \cite{SupMat}. In contrast, the ground state we obtain via imaginary time evolution is a single droplet for the full range of $\lambda_{xy}$ values.
    Within the assumptions of our theoretical analysis and its comparison with the experiment, we conclude that although in a more controlled environment, the condensate undergoes a modulational instability similar to the one observed in \cite{Kadau2016}. This instability stems from the joint effect of the trapping potential and the dipole-dipole interaction. It is related, in the case of large, confined BECs, to the instability due to roton-softening recently observed in \cite{Chomaz2017}. For the first time we induce such a modulational instability by controlling the DDI rather than the scattering length.
    
    As a consequence, the striped states we observe are likely not the ground state predicted from our study of the Gross-Pitaevskii equation, but rather an excited metastable state. We further emphasize that the ``roton'' modes driving the instability of the initial BEC in general do not lead to the ground state of the system. Instead, the excited state observed in the experiment is prevented from dissipating its energy and reaching the ground state \cite{Wachtler2016}. This can be understood as the consequence of a strong energy barrier existing between two droplets, see Fig. \ref{fig:S1}a). This barrier also prevents the preparation of the ground state by first preparing a single droplet and then compressing it.

    \section{Phase Coherence}

    \begin{figure}
        \begin{overpic}[width=0.45\textwidth]{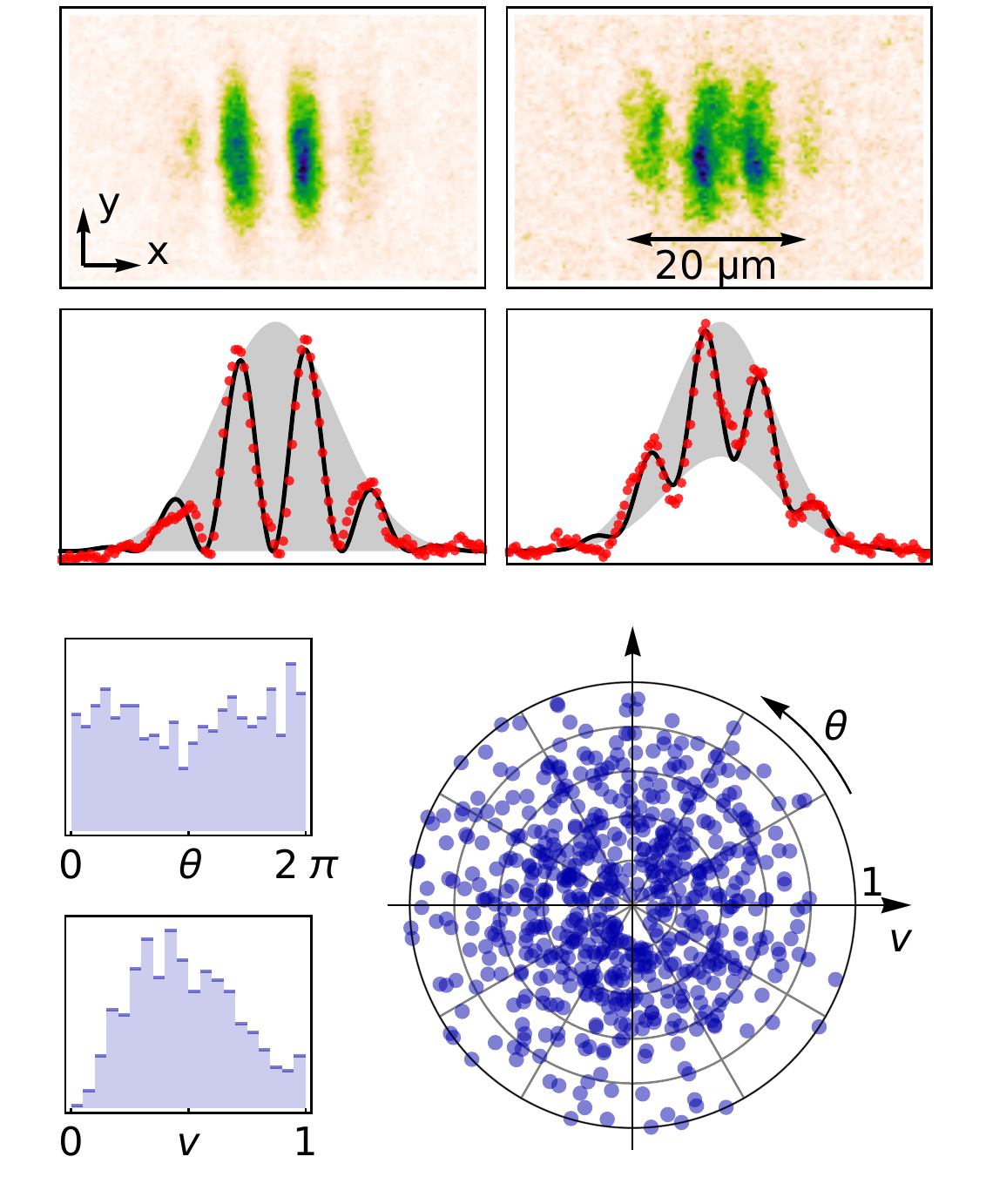}
            \put(0,96){a)} \put(0,44){b)}
        \end{overpic}\vspace{-3mm}
        \caption{\label{fig:4} Interference patterns after $8\unit{ms}$ of expansion.
            a) Two example realizations with absorption image (top) and integrated density (bottom) showing fringes. We extract the phase $\theta$ with respect to the center of mass of the distribution.
            b) Histograms and polar plot of relative phase $\theta$ and visibility $v$ for 650 atom distributions. There is no preferred phase visible indicating that there is no phase coherence between the droplets.}
    \end{figure}
    
    Although not the ground state, a metastable striped state might still feature phase coherence between the droplets, and thus make it a candidate for a metastable supersolid state of matter. To explore this possibility experimentally we conduct interference experiments. In prior experiments \cite{Ferrier-Barbut2016} fringe patterns have been observed indicating local superfluidity of the quantum droplets. Here, we focus on the phase relation between the droplets.
    As known from the physics of condensates in double-well potentials \cite{Gati2007} the expansion of two wavepackets with phases $\theta_1$ and $\theta_2$ results in interference peaks with a phase $\theta = \theta_1 - \theta_2$ relative to the envelope. Figure \ref{fig:4}a) shows example realizations of such fringe patterns in our experiment obtained via absorption imaging after $8\unit{ms}$ of expansion with the magnetic field $B = 1245(5)\unit{mG}$ ramped to $B = 1313(5)\unit{mG}$ during the first $2\unit{ms}$. After integration along the $y$ direction we extract the relative phase $\theta$ and fringe visibility $v$ with a cosine-modulated Gaussian function, see \cite{SupMat}.
    The extracted values $(v,\theta)$ for 650 realizations in a $\lambda_{xy} = 1/4$ trap (3-4 trapped droplets initially) with subsequent expansion are plotted in Fig{.} \ref{fig:4}b). For a phase-coherent sample a fixed phase relation with a single predominant value of $\theta$ is expected. Yet the experiment shows a random phase distribution and thus no phase-coherence between the droplets. We note that experiments with different starting conditions, especially varying the initial droplet number including pairs of droplets, show the same behavior.
    
    The mentioned finite-wavelength instability causing the initial BEC to split up into multiple droplets \cite{Kadau2016} is stochastic in nature. This induces inevitable atom number fluctuations in the droplets and additional phase noise. An atom number difference between interfering droplets leads to a difference in chemical potential, which is known to cause a random relative dephasing of the droplets during the preparation time.
    
    \begin{figure}
        \begin{overpic}[width=0.45\textwidth]{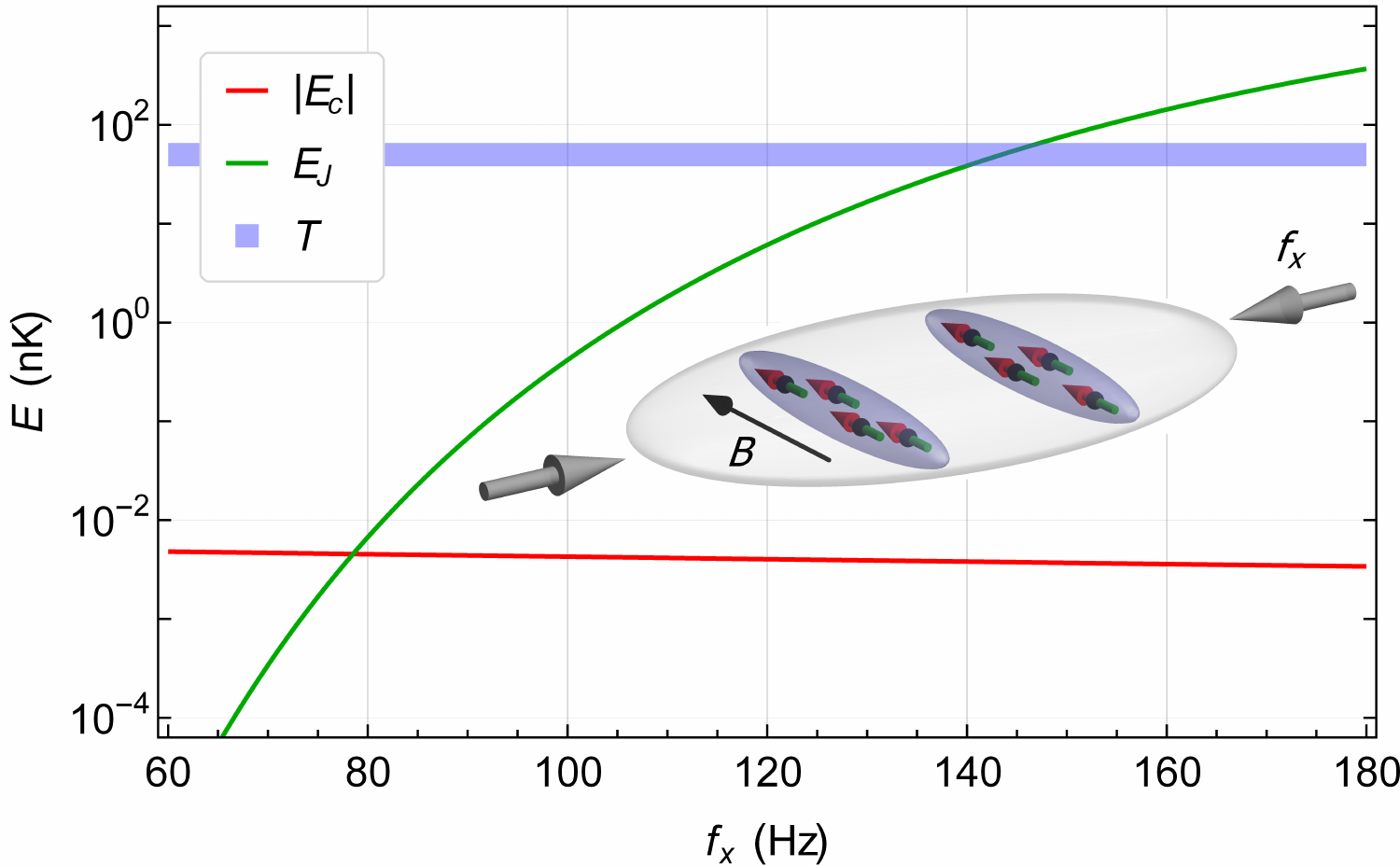}\end{overpic} \vspace{-3mm}
        \caption{\label{fig:5} Tuning the coherence properties. On-site interaction $E_C$ (red) and hopping term $E_J$ (green line) as defined in \cite{SupMat} for varying confinement $f_x$ perpendicular to the magnetic field.  This way the distance between droplets is varied and thus $E_J$ can be tuned over several orders of magnitude. The shape of the droplet wavefunction is not altered keeping $E_C$ almost constant. The values are computed with the variational ansatz for a confinement where the double-droplet solution is the ground state ($f_y = 300\unit{Hz}$). The temperature $T = 30 - 70 \unit{nK}$ reflects the temperature range in the experiment.}
    \end{figure}
    
    While our experiments show no sign of phase coherence, we present a mechanism that could establish phase coherence in future experiments. Therefore we resort to the well-developed framework of bosonic Josephson junctions \cite{Smerzi1997,Gati2007}. Using our double-droplet variational ansatz we can develop a two-state model in direct analogy to \cite{Smerzi1997}, which was also studied in \cite{Abad2011} for self-induced bosonic Josephson junctions in dipolar BECs. This yields the on-site interaction term $E_C$ in first approximation, which contains the contact and dipolar mean-field on-site values as well as the dipolar inter-site mean-field and a beyond-mean-field on-site term. In addition we can calculate the usual tunneling term $E_J$, as detailed in \cite{SupMat}. This model is by no means an analysis of quantum and thermal fluctuations in this system, but it provides an estimate of the parameter range where the phase links between droplets should be robust against these effects. The criteria for robustness against phase fluctuations due to quantum noise in a bosonic Josephson junction \cite{Gati2007} is given by $| E_J / E_C | > 1$, while against thermal fluctuations this is $E_J / k_B T > 1$.
    
    In our system, owing to the attractive dipolar interaction within the droplet, $E_C$ is negative with values of $|E_C|\ll 1\unit{nK}$. For the experimental parameters shown here the hopping term is $E_J \ll 1\unit{nK}$ as well.
    Yet, since the hopping term is a measure of the droplet wavefunction overlap it scales exponentially with the distance between the droplets. Therefore, the hopping energy can be tuned over several orders of magnitude by varying $f_x$, the confinement perpendicular to the droplet axis, see Fig{.} \ref{fig:5}. This primarily changes the droplet distance preserving the droplet sizes and thus $E_C$. For the largest confinement considered we obtain values of $E_J > 200\unit{nK}$ that are distinctively larger than the temperature range $T \approx 30-70\unit{nK}$ observed in the experiment. In a setting with stronger confinement we thus expect to drive the multiple droplet states to the regime $E_J/k_B T > 1$, where thermal fluctuations do not prohibit phase coherence \cite{Gati2006}. 
    However, the role of the dephasing mechanisms needs to be investigated first. This includes a systematic study of the atom number fluctuations in samples of multiple droplets, which is the topic of future work.\\

    \section{Conclusion}

    To summarize, we predict a striped ground state for a dipolar Bose gas in a harmonic trap. A variational ansatz for two dipolar quantum droplets shows a transition from a single- to a double-droplet state for increasing confinement along the droplet's long axis. Numerical simulations of the eGPE support this observation and reveal transitions to states with higher droplet numbers for larger confinement. Within the Gross-Pitaevskii theory this is a supersolid state at zero temperature.
    
    In the experiment, we induce the transition from the BEC to the self-organized phase by tilting the magnetic field in an anisotropic trap. The $z$ confinement as well as the scattering length influence the value of the respective critical tilt angle $\beta_c$, since they alter the dipolar mean-field interaction potential. By reshaping the trap we further observe striped states with multiple droplets in situ. Because of the modulational instability causing the fragmentation of the initial BEC, we observe higher droplet numbers in the experiment compared to the single-droplet ground state expected from theory. These states are thus metastable density-modulated states, that cannot decay to the expected ground state since there is an energy barrier. However, the observed striped states are long-lived and could share the coherence properties of the ground state. As such these states are accessible in experiments, but have not yet been investigated theoretically. 
    
    The stochastic nature of the droplet creation leads to number fluctuations and thus the loss of a common phase relation. Although we do not observe signs of mutual phase coherence between the droplets in interference experiments, we point out a mechanism that might establish phase coherence throughout the sample.
    
    In this work we have established the framework and necessary tools to characterize a supersolid phase of a dipolar Bose gas. Future work will be directed towards understanding the dephasing mechanism to finally realize a dipolar supersolid state of matter.\\ \vspace{3mm}

    \begin{acknowledgments}
        We thank A{.} Smerzi for valuable discussions. This work is supported by the German Research Foundation (DFG) within SFB/TRR21 and FOR2247. IFB and TL acknowledge support from the EU within Horizon2020 Marie Sk{\l}odowska Curie IF (703419 DipInQuantum and 746525 coolDips, respectively). TL acknowledges support from the Alexander von Humboldt Foundation through a Feodor Lynen Fellowship.
    \end{acknowledgments}

    \appendix

    \section{Theory}

        \subsection{Variational Ansatz}
\begin{figure}
            \begin{overpic}[width=0.45\textwidth]{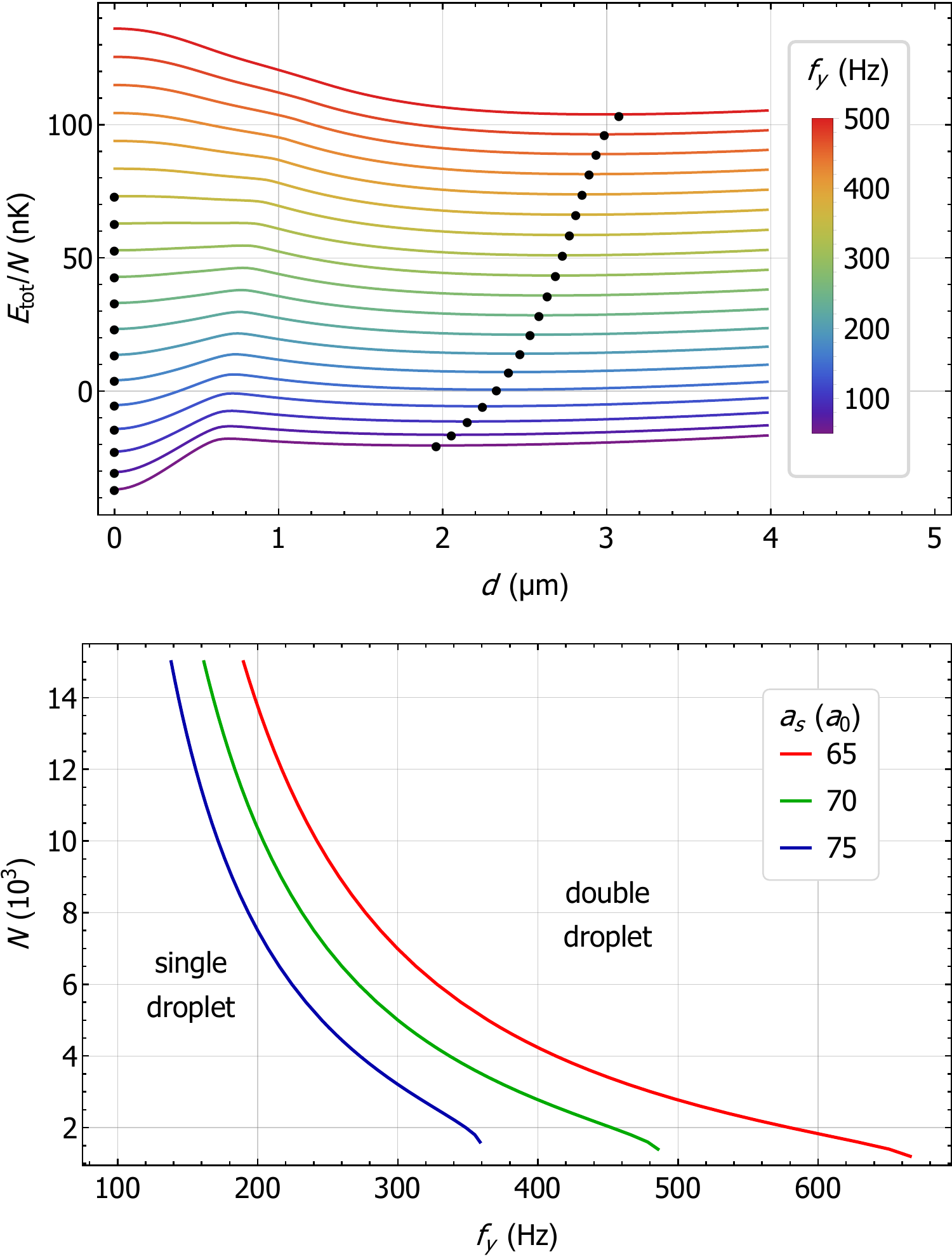}
                \put(0,97){a)} \put(0,47){b)} \end{overpic} \vspace{-2mm}
            \caption{\label{fig:S1} Calculations with the variational ansatz. a) Energy per particle $E_\mathrm{tot}/N$ over distance $d$ acquired with the variational ansatz. Curves represent calculations for different trap frequencies $f_y = 2 \pi \cdot 50 - 500 \unit{Hz}$. The solution for a single droplet ($d = 0$) and for a double droplet state with distance $d > 0$ are marked.
                b) Phase diagram of atom number $N$ over confinement $f_x$ along the droplet axis.
                Solid lines mark the transition between the single-droplet ground state (left) and the double-droplet ground state (right) for different values of the scattering length $a_s$.
                See main text for parameters.}
        \end{figure}

        We extend the ansatz for a single droplet \cite{Baillie2016,Wachtler2016a} with a gaussian trial wavefunction
        \begin{equation}\label{eq:Gauss}
        \psi_0(x,y,z) = \sqrt{\frac{N_d}{\pi^{3/2} \sigma_x\sigma_y\sigma_z }} 
        \, e^{ -\frac{1}{2} \!\left( \frac{x^2}{\sigma_x^2} + \frac{y^2}{\sigma_y^2} + \frac{z^2}{\sigma_z^2} \right) }
        \end{equation}
        to the case of two droplets at position $x=\pm d/2$. Curves in Fig{.} 1a) are obtained by minimizing the energy functional $E_{tot}\big[|\psi_0(x-d/2,y,z)|^2+|\psi_0(x+d/2,y,z)|^2\big]$ corresponding to the eGPE of eq. (1) with respect to the sizes $\sigma_k$ and droplet distance $d$. We calculate the energy along the lines of \cite{Giovanazzi2006a,Glaum2007} and obtain the single contributions
        \begin{eqnarray}\label{eq:Evar}
        \frac{E_\mathrm{kin}}{N \hbar\bar\omega} &=& \frac{\bar{a}^2}{4} \Big( \sigma_x^{-2} + \sigma_y^{-2} + \sigma_z^{-2} \Big) \nonumber\\
        \frac{E_\mathrm{pot}}{N \hbar\bar\omega} &=& \frac{1}{4 \bar{a}^2} \left[ \omega_x^2\Big(\sigma_x^{2} + \frac{1}{2}d^2\Big) + \omega_y^2\sigma_y^{2} + \omega_z^2\sigma_z^{2} \right] \nonumber\\
        \frac{E_\mathrm{con}}{N \hbar\bar\omega} &=& \frac{\bar{a}^2}{2 \sqrt{2\pi}\bar{\sigma}^3} N a_s \Big( 1 + e^{-\frac{1}{2}u^2} \Big) \nonumber\\
        \frac{E_\mathrm{dip}}{N \hbar\bar\omega} &=& \frac{\bar{a}^2}{2 \sqrt{2\pi}\bar{\sigma}^3} N a_{dd} \Big( \!-f(\kappa_x, \kappa_y) - \mathcal{I}_\mathrm{dip}( \kappa_x, \kappa_y, u ) \Big) \nonumber\\
        \frac{E_\mathrm{qf}}{N \hbar\bar\omega} &=& \frac{512 \sqrt{2} \bar{a}^2}{75 \sqrt{5} \pi^{7/4} \bar{\sigma}^{9/2}} N^{3/2} a_s^{5/2} \Big( 1 + \frac{3}{2}\frac{a_{dd}^2}{a_s^2} \Big) \,\mathcal{I}_\mathrm{qf}(u) \quad\quad
        \end{eqnarray}
        to the total energy $E_\mathrm{tot}$
        with aspect ratios $\kappa_k = \sigma_k/\sigma_z$, rescaled distance $u = d/\sigma_x$ as well as mean size $\bar\sigma = (\sigma_x \sigma_y \sigma_z)^{1/3}$ and mean harmonic oscillator length $\bar a = \sqrt{\hbar m/{ (\omega_x \omega_y \omega_z)^{1/3}}}$ introduced here. The integrals 
        \begin{eqnarray}
        &\mathcal{I}_\mathrm{dip}&(\kappa_x, \kappa_y, u)
        = e^{-\frac{1}{2}u^2} - 3 \frac{\kappa_x \kappa_y}{(1-\kappa_x^2)^{3/2}} \cdot \nonumber \\
        && \int_0^{\sqrt{1-\kappa_x^2}}\!\mathrm{d}\xi \frac{\xi^2 \exp\!\left(-\frac{u^2}{2} \frac{\kappa_x^2 \xi^2}{(1-\kappa_x^2)(1-\xi^2)} \right) }{\sqrt{1-\xi^2} \sqrt{1-\xi^2 \frac{1-\kappa_y^2}{1-\kappa_x^2} }}
        \end{eqnarray}
        with $f(\kappa_x, \kappa_y ) = \mathcal{I}_\mathrm{dip}(\kappa_x, \kappa_y, 0)$ for the dipolar interaction and
        \begin{equation}
        \mathcal{I}_\mathrm{qf}(u) = \frac{2}{\sqrt{\pi}} e^{-\frac{5}{8} u^2} \int_0^\infty\!\mathrm{d}v \, e^{-v^2} \!\cosh\!\left( \sqrt{\frac{2}{5}} u v\right)^{5/2} 
        \end{equation}
        for the quantum fluctuations stem from the overlap of the two wavefunctions. The solution for a single droplet is recovered for $d=0$.
        For the parameter range discussed in the main text we find local minima of the energy functional at both $d = 0$ and $d > \sigma_x > 0$ corresponding to the solutions for a single and double droplet state, see black dots in Fig{.} \ref{fig:S1}a).
        By definition the variational ansatz overestimates the energy.
        
        \subsection{Coherence Properties}
        We estimate the on-site interaction energy $E_C$ and hopping term $E_J$ within the framework of BECs in double-well potentials \cite{Smerzi1997,Abad2011,Gati2007}. Within the two-state model we use the single-droplet wavefunction 
        \begin{equation}\label{eq:GaussOne}
        \phi_{1,2} = \frac{1}{\pi^{3/4} \bar{\sigma}^{3/2}} \exp\left(-\frac{(x\pm d/2)^2}{2\sigma_x^2} -\frac{y^2}{2\sigma_y^2} -\frac{z^2}{2 \sigma_z^2}\right)
        \end{equation}
        normalized to $\int\!\mathrm{d}^3r \left| \phi_{1,2} \right|^2 = 1$ and extend the on-site interaction
        \begin{eqnarray}\label{eq:EC}
        E_C &=& \int\!\!\mathrm{d}^3r\, \!\bigg[ g \left|\phi_{1}\right|^4 + g_{qf} \sqrt{\frac{N}{2}} \left|\phi_{1}\right|^5 \nonumber \\
        &&+ \left( \left|\phi_{1}\right|^2 + \left|\phi_{2}\right|^2 \right) \!\!\int\!\!\mathrm{d}^3r^\prime V_d \left|\phi_{1}^\prime\right|^2\bigg] \quad \nonumber \\
        &=& g \frac{1}{(2\pi)^{3/2}\bar{\sigma}^3} + g_{qf} \frac{2 N^{1/2}}{5^{3/2}\pi^{9/4}\bar{\sigma}^{9/2}} \nonumber \\
        &&- g_{dd} \frac{f(\kappa_x,\kappa_y) + \mathcal{I}_\mathrm{dip}(\kappa_x, \kappa_y, u) 
        }{(2\pi)^{3/2}\bar{\sigma}^3}
        \end{eqnarray}
        by quantum fluctuations as well as the inter-droplet repulsion of the dipolar interaction. We note that $E_C$ can become negative because of the attractive dipolar interaction in a stable quantum droplet.
        The hopping term
        \begin{eqnarray}\label{eq:EJ}
        E_J &=& N \int\!\!\mathrm{d}^3r\, \phi_1 \!\left( -\frac{\hbar^2 \nabla^2}{2 m} + V_{ext} \right)\! \phi_2 \nonumber \\
        &=& - \frac{\hbar^2}{4 m} e^{-\frac{u^2}{4}} \!\left[ \sum_{k=x,y,z} \!\!\!\left(\sigma_k^{-2}  + \frac{m^2 \omega_k^2}{\hbar^2} \sigma_k^{2}\right) - \frac{u^2}{2 \sigma_x^2}\right] \quad\quad
        \end{eqnarray}
        only depends on quantum pressure and the external potential. As a measure for the wavefunction overlap it scales exponentially with the rescaled distance $u$ between droplets. Thus $E_J$ can be tuned over a wide range, as demonstrated in Fig{.} 5) of the main text.
        
        \subsection{Dynamic simulations}
        As in the experiment, a BEC is prepared at an magnetic field angle $\beta \ll \beta_c$ via imaginary-time evolution of the eGPE. Subsequently, the angle is tilted at a constant speed of $\dot\beta = 0.33\unit{deg/ms}$ in real-time evolution. At the critical angle $\beta_c$ we observe the transition to one or multiple droplets depending on the transversal trap aspect ratio $\lambda_{xy}$. A marker for this transition is the combined two-body energy $E_{con} \!+\! E_{dip}$ that becomes negative for $\beta \ge \beta_c$. The overall droplet number in the simulation is slightly higher compared to the experiment and crucially depends on both atom number and scattering length where we chose values of $N = 5000$ and $a_s = 70\,a_0$ close to the experiment. Yet the former is subject to experimental fluctuations and the latter is not known very precisely. 
        We also take into account three-body losses by an additional term $-i \frac{\hbar}{2} L_3 |\psi|^4$ in the eGPE with loss constant $L_3 = 1.25 \cdot 10^{-41}\unit{m^6\!/s}$ \cite{Schmitt2016a}. Due to atom loss of the single droplets two of these can merge into a single one, lowering the droplet atom number for $\beta \gg \beta_c$.

    \section{Experiment}

        \subsection{Setup}
        Our apparatus creates Bose-Einstein condensates of the isotope $^{164}\mathrm{Dy}$ in a crossed optical dipole trap (along $\hat x$ and $\hat y$ axes, $\lambda = 1064\unit{nm}$) with a microscope objective allowing for in situ imaging ($1\unit{\mu m}$ resolution) along the $\hat z$ axis. The light sheet is implemented diagonally in the imaging plane parallel to the $(\hat x +\hat y)$ axis. By reshaping a round gaussian beam at $\lambda = 532\unit{nm}$ with cylindrical lenses we obtain a waist $w_z\approx 4\unit{\mu m}$ leading to measured trap frequencies of $f_z \le 2.0\unit{kHz}$ for the strong axis.
        In the experimental cycle we ramp up the light sheet power in $50\unit{ms}$ after generating a Bose-Einstein condensate in the crossed infrared trap. We use the infrared beams to simultaneously reshape the trap in the $xy$ plane. Afterwards we tilt the magnetic field with constant rate $\dot\beta$ and magnitude $B$. Then we typically wait for $10\unit{ms}$ and finally use phase-contrast imaging at a detuning of $10\,\Gamma$ to measure the density distribution integrated along the $z$ axis in situ.

        \subsection{Fourier Anisotropy}
        In order to quantify the transition from the condensate to the droplet phase at a critical angle $\beta_c$ we analyze the Fourier transform of the acquired images. For each image $I$ we compute the spectrum $|\mathcal{F}(I)|^2$ and sum it over an area of width $\Delta k_{y (x)} = 4\unit{\mu m^{-1}}$ along the x (y) axis, respectively. The difference of these sums normalized to the sum over the combined area defines the fourier anisotropy $A_{FT}$. Relying on the anisotropy of the cloud's aspect ratio, this quantity is independent of the observed droplet number. The $A_{FT}(\beta)$ data is typically averaged over 4-10 realizations.
        In order to extract the critical angle $\beta_c$ we use the empirical fit function $A_{FT}(\beta) \propto \arctan\!\left((\beta - \beta_c)/w\right)$, as shown in Fig{.} 2b). The error is given by the quadratic mean of the fit error and the width $w$.

        \subsection{Tilt Speed}
\begin{figure}[h]
            \begin{overpic}[width=0.45\textwidth]{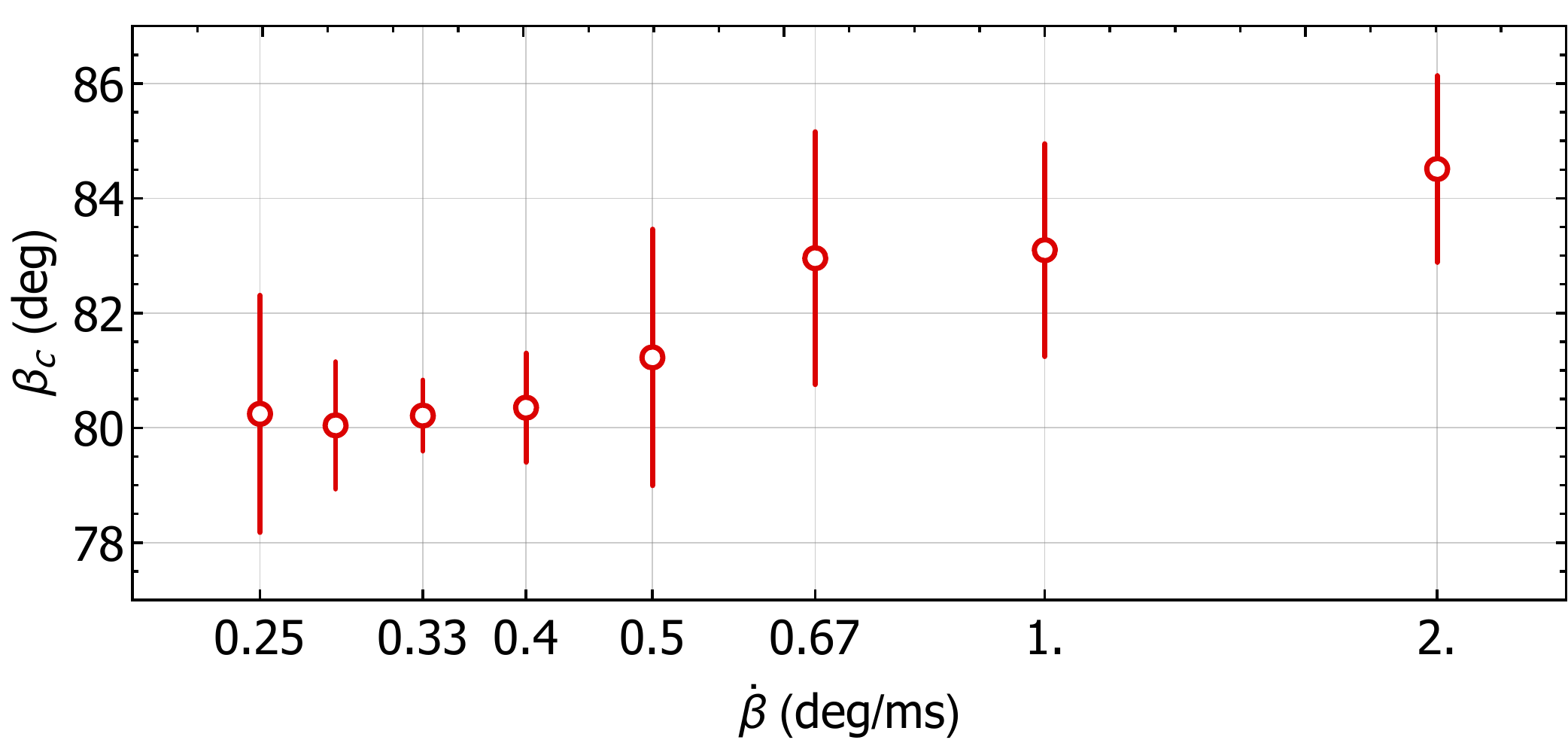}
            \end{overpic}\vspace{-3mm}
            \caption{\label{fig:S2} Measured critical angle $\beta_c$ over tilt speed $\dot\beta$.}
        \end{figure}

        We observe a larger critical angle when tilting the magnetic field faster than $\dot \beta \le 0.4\unit{deg/ms}$, see Fig{.} \ref{fig:S2}). This is due to the finite time of the collapse dynamics.
        The creation time of the droplets was measured to be $\approx 15\unit{ms}$ (comparable to $\approx7\unit{ms}$ after a field quench in a different geometry \cite{Kadau2016}).
        Data presented in this article is thus taken at a relatively slow tilt speed of $\dot \beta = 0.33\unit{deg/ms}$, where the critical angle is not overestimated.

        \subsection{Interference Patterns}
        For the interference experiments we record the interference patterns via absorption imaging after free expansion. We integrate the images along $y$ to obtain the integrated density $n_\mathrm{int}(x)$, see Fig{.} 4a). Fitting the cosine-modulated gaussian function
        \begin{equation}\label{eq:FringeFit}
        n_\mathrm{int}(x) \propto e^{-\frac{(x-x_0)^2}{2\sigma^2}} \!\left[ 1 + v \cos( k (x-x_0) + \theta) \right]
        \end{equation}
        to the data allows to extract the phase $\theta$ with respect to the center of mass position $x_0$ of the distribution as well as the visibility $0 \leq v \leq 1$ of fringes. 
        These two quantities are shown in Fig{.} 4b) in polar coordinates $(v,\theta$).
        We note that the extracted gaussian size $\sigma$ and wavelength $\lambda = 2\pi/k$ have the same magnitude and thus only a few fringes are visible.

    
    \bibliography{paper}

\end{document}